\documentstyle[12pt]{article}
\input epsf.tex

\begin{document}

\baselineskip=15.5pt
\pagestyle{plain}
\setcounter{page}{1}

\renewcommand{\thefootnote}{\fnsymbol{footnote}}

\begin{titlepage}

\begin{flushright}
PUPT-1954\\
hep-th/0009139
\end{flushright}
\vfil

\begin{center}
{\huge TASI Lectures:
 Introduction to the \\} 
 \vspace{3 mm}
{\huge AdS/CFT Correspondence
}

\end{center}

\vfil

\begin{center}
{\large Igor R.\ Klebanov}\\
\vspace{3mm}
Joseph Henry Laboratories\\
Princeton University\\
Princeton, New Jersey 08544, USA\\
\vspace{3mm}
\end{center}

\vfil

\begin{center}
{\large Abstract}
\end{center}

\noindent
This is an introductory review of the AdS/CFT correspondence
and of the ideas that led to its formulation. We show how
comparison of stacks of D3-branes with corresponding supergravity
solutions leads
to dualities between conformal large $N$ gauge theories in
4 dimensions and string backgrounds of the form $AdS_5\times X_5$
where $X_5$ is an Einstein manifold. 
The gauge invariant chiral operators of the field theory
are in one-to-one correspondence with the supergravity modes,
and their correlation functions at strong `t Hooft coupling are
determined by the dependence of the supergravity action on AdS
boundary conditions. The simplest case
is when $X_5$ is a 5-sphere and the dual gauge theory is the
${\cal N}=4$ supersymmetric $SU(N)$ Yang-Mills theory. We also discuss
D3-branes on the conifold corresponding
to $X_5$ being a coset space $T^{1,1}=(SU(2)\times SU(2))/U(1)$.
This background is dual to a certain ${\cal N}=1$ superconformal field theory 
with gauge group $SU(N)\times SU(N)$.
\vfil
\begin{flushleft}
9/00
\end{flushleft}
\end{titlepage}
\newpage

\renewcommand{\thefootnote}{\arabic{footnote}}
\setcounter{footnote}{0}


\newcommand{\grad}{\nabla}
\newcommand{\tr}{\mathop{\rm tr}}
\newcommand{\half}{{1\over 2}}
\newcommand{\third}{{1\over 3}}
\newcommand{\be}{\begin{equation}}
\newcommand{\ee}{\end{equation}}
\newcommand{\bea}{\begin{eqnarray}}
\newcommand{\eea}{\end{eqnarray}}

\newcommand{\dint}[2]{\int\limits_{#1}^{#2}}
\newcommand{\D}{\displaystyle}
\newcommand{\PDT}[1]{\frac{\partial #1}{\partial t}}
\newcommand{\PD}{\partial}
\newcommand{\tw}{\tilde{w}}
\newcommand{\tg}{\tilde{g}}
\newcommand{\newcaption}[1]{\centerline{\parbox{6in}{\caption{#1}}}}
\def\href#1#2{#2}  
\def\Z{{\bf Z}}
\def\N{{\bf N}}
\def\Tr{{\rm Tr}}
\def \ci {\cite}
\def \foot {\footnote}
\def \bi{\bibitem}
\newcommand{\rf}[1]{(\ref{#1})}
\def \del{\partial}
\def \m {\mu}
\def \n {\nu} 
\def \g {\gamma}
\def \G {\Gamma}
\def \a {\alpha}
\def \ov {\over}
\def \la {\label}
\def \ep {\epsilon}
\def \d {\delta}
\def \k {\kappa}
\def \p {\phi}
\def \ha {\textstyle{1\ov 2}}

\def\np {  {\em Nucl. Phys.} }
\def \pl { {\em Phys. Lett.} }
\def \mpl { Mod. Phys. Lett. }
\def \prl { Phys. Rev. Lett. }
\def \pr  { {\em Phys. Rev.} }
\def \cqg { Class. Quantum Grav.}
\def \jmp { Journ. Math. Phys. }
\def\ap { Ann. Phys. }
\def \ijmp { Int. J. Mod. Phys. }

%
\def\TL{\hfil$\displaystyle{##}$}
\def\TR{$\displaystyle{{}##}$\hfil}
\def\TC{\hfil$\displaystyle{##}$\hfil}
\def\TT{\hbox{##}}
\def\seqalign#1#2{\vcenter{\openup1\jot
  \halign{\strut #1\cr #2 \cr}}}

\def\comment#1{}
\def\fixit#1{}

\def\tf#1#2{{\textstyle{#1 \over #2}}}
\def\df#1#2{{\displaystyle{#1 \over #2}}}

\def\mop#1{\mathop{\rm #1}\nolimits}

\def\ad{\mop{ad}}
\def\coth{\mop{coth}}
\def\csch{\mop{csch}}
\def\sech{\mop{sech}}
\def\Vol{\mop{Vol}}
\def\vol{\mop{vol}}
\def\diag{\mop{diag}}
\def\tr{\mop{tr}}
\def\Disc{\mop{Disc}}
\def\sgn{\mop{sgn}}

\def\SU{{\rm SU}}
\def\USp{{\rm USp}}            

\def\lsim{\mathrel{\mathstrut\smash{\ooalign{\raise2.5pt\hbox{$<$}\cr\lower2.5pt\hbox{$\sim$}}}}}
\def\gsim{\mathrel{\mathstrut\smash{\ooalign{\raise2.5pt\hbox{$>$}\cr\lower2.5pt\hbox{$\sim$}}}}}

\def\slashed#1{\ooalign{\hfil\hfil/\hfil\cr $#1$}}

\def\sqr#1#2{{\vcenter{\vbox{\hrule height.#2pt
         \hbox{\vrule width.#2pt height#1pt \kern#1pt
            \vrule width.#2pt}
         \hrule height.#2pt}}}}
\def\square{\mathop{\mathchoice\sqr56\sqr56\sqr{3.75}4\sqr34\,}\nolimits}

\def\idget{$\sqr55$\hskip-0.5pt}
\def\endrow{\hskip0.5pt\cr\noalign{\vskip-1.5pt}}
\def\endyoung{\hskip0.5pt\cr}

\def\href#1#2{#2}  


%
\def\lbldef#1#2{\expandafter\gdef\csname #1\endcsname {#2}}
\def\eqn#1#2{\lbldef{#1}{(\ref{#1})}%
\begin{equation} #2 \label{#1} \end{equation}}
\def\eqalign#1{\vcenter{\openup1\jot
    \halign{\strut\span\TL & \span\TR\cr #1 \cr
   }}}
\def\eno#1{(\ref{#1})}

\def\rmax{{r_{\rm max}}}
\def\gone#1{}

\section{Introduction}

String theory originated from attempts to understand 
the strong interactions \cite{NNS}. However, after the emergence of QCD
as the theory of hadrons, the dominant theme of string research
shifted to the Planck scale domain of quantum gravity \cite{Scherk}. 
Although in hadron physics one routinely hears about flux tubes
and the string tension, many particle theorists gave up hope
that string theory might lead to an exact description 
of the strong interactions.
Now, however, due to the progress that has taken place
over the last 3 years, we can say with confidence
that at least some strongly coupled gauge theories have a dual description
in terms of strings. Let me emphasize that one is not talking here about
effective strings that give an approximate qualitative description
of the QCD flux tubes, but
rather about an exact duality. At weak gauge coupling a convenient
description of the theory involves conventional perturbative methods;
at strong coupling, where such methods are intractable, the dual string
description simplifies and gives exact information about the theory.
The best established examples of this duality are 
(super)conformal gauge theories
where the so-called AdS/CFT correspondence \cite{jthroat,US,EW}
has allowed for many calculations
at strong `t Hooft coupling to be performed with ease. 
In these notes I describe,
from my own personal perspective, some of the ideas that led to the formulation
of the AdS/CFT correspondence, as well as more recent results. 
For the sake of brevity I will mainly discuss the
AdS$_5$/CFT$_4$ case which is most directly related to 4-dimensional
gauge theories.

It has long been believed that the best hope for a string description
of non-Abelian gauge theories lies in the 't~Hooft large $N$ limit.
A quarter of a century ago 't~Hooft proposed to generalize the $SU(3)$
gauge group of QCD to $SU(N)$, and to take the large $N$ limit while keeping
$g_{{\rm YM}}^2 N$ fixed \cite{GT}. In this limit each Feynman graph carries
a topological factor $N^\chi$, where $\chi$ is the Euler
characteristic of the graph.
Thus, the sum over graphs of a given topology can perhaps be thought of
as a sum over world sheets of a hypothetical ``QCD string.'' 
Since the spheres (string tree diagrams)
are weighted by $N^2$, the tori (string one-loop diagrams) -- 
by $N^0$, etc., we find that the closed string coupling constant is of order
$N^{-1}$. Thus, the advantage of taking $N$ to be large is that we find
a weakly coupled string theory. It is not clear, however, how to describe
this string theory in elementary terms (by a 2-dimensional world sheet action, 
for example). This is clearly an important problem because the 
spectrum  of such free closed strings is just the  
spectrum of glueballs in large $N$ QCD.
If quarks are included, then we also find open strings describing the 
mesons. Thus, if methods are developed for calculating these spectra,
and it is found that they are discrete, then
this provides an elegant
explanation of confinement. Furthermore, the $1/N$ corrections 
correspond to perturbative string corrections.

Many years of effort, and many good ideas, were invested into the search
for an exact gauge field/string duality \cite{book}. 
One class of ideas, exploiting
the similarity of the large $N$ loop equation with the string Schroedinger
equation, eventually led to the following fascinating 
speculation \cite{Sasha}:
one should not look for the QCD string in four dimensions, but rather
in five, with the fifth dimension akin to the Liouville dimension of
non-critical string theory \cite{bos}. This leads to a picture
where the QCD string is described by a two-dimensional world sheet
sigma model with a curved non-compact 5-dimensional target space. 
The difficult question is:
precisely which target spaces are relevant to gauge theories?
Luckily, we now do have answers to this question for a variety of
conformal large $N$ gauge models. In these
examples of the gauge field/string duality the strings
propagate in 5 compact dimensions in addition to the 5 non-compact
ones. In fact, these ``gauge strings'' are none other than type IIB
superstrings propagating in curved 10-dimensional backgrounds of the
form $AdS_5\times X_5$. The $AdS_5$ factor
present in the dual description of all conformal
field theories is the 5-dimensional Anti-de Sitter
space which has constant negative curvature. $X_5$ is a compact 
positively curved Einstein space which depends on the specific model: the
simplest example is when $X_5$ is a 5-sphere leading to the dual
formulation of the ${\cal N}=4$ supersymmetric Yang-Mills theory
\cite{jthroat,US,EW} but
other, more intricate, dualities have also been constructed
\cite{ks,lnv,KW}.
The route that leads to these results 
involves an unexpected detour via black holes and Dirichlet branes.
We turn to these subjects next.

\section{D-branes vs. Black Holes and $p$-branes}

A few years ago it became clear that, 
in addition to strings, superstring theory
contains soliton-like ``membranes'' of various internal dimensionalities
called Dirichlet branes (or D-branes) \cite{Dnotes}.
A Dirichlet $p$-brane (or D$p$-brane) is a $p+1$ dimensional hyperplane
in $9+1$ dimensional space-time where strings are allowed to end,
even in theories where all strings are closed in the bulk of space-time.
In some ways a D-brane is like a topological defect: when a closed string
touches it, it can open up and turn into an open string whose
ends are free to move along the D-brane. For the end-points of such a string
the $p+1$ longitudinal coordinates satisfy the conventional free (Neumann)
boundary conditions, while the $9-p$ coordinates transverse to
the D$p$-brane have the fixed (Dirichlet) boundary conditions; hence
the origin of the term ``Dirichlet brane.'' 
In a seminal paper \cite{brane} Polchinski
showed that the D$p$-brane is a BPS saturated object which preserves
$1/2$ of the bulk supersymmetries and carries an elementary unit
of charge with respect to the $p+1$ form gauge potential from the
Ramond-Ramond sector of type II superstring. The existence of BPS
objects carrying such charges is required by non-perturbative string
dualities \cite{HullT}. A striking feature of the D-brane formalism 
is that it provides a
concrete (and very simple) embedding of such objects into perturbative
string theory.

Another fascinating feature of the D-branes is that they naturally realize
gauge theories on their world volume. The massless spectrum of open strings
living on a D$p$-brane is that of a maximally supersymmetric $U(1)$
gauge theory in $p+1$ dimensions. The $9-p$ massless
scalar fields present in this supermultiplet are the expected Goldstone 
modes associated with the transverse oscillations of the D$p$-brane,
while the photons and fermions
may be thought of as providing the unique supersymmetric
completion.
If we consider $N$ parallel D-branes,
then there are $N^2$ different species of open strings because they can
begin and end on any of the D-branes. 
$N^2$ is the dimension of the adjoint representation of $U(N)$,
and indeed we find the maximally supersymmetric $U(N)$ 
gauge theory in this setting \cite{Witten}.
The relative separations of the D$p$-branes in the $9-p$ transverse
dimensions are determined by the expectation values of the scalar fields.
We will be primarily interested in the case where all scalar expectation
values vanish, so that the $N$ D$p$-branes are stacked on top of each other.
If $N$ is large, then this stack is a heavy object embedded into a theory
of closed strings which contains gravity. Naturally, this macroscopic
object will curve space: it may be described by some classical metric
and other background fields including the Ramond-Ramond 
$p+1$ form potential.
Thus, we have two very different descriptions of the stack of D$p$-branes:
one in terms of the $U(N)$ supersymmetric gauge theory on its world volume,
and the other in terms of the classical
Ramond-Ramond charged $p$-brane background of the type II
closed superstring theory. The relation between these two descriptions
is at the heart of the recent progress in understanding 
connections between gauge fields and strings that are the subject of
these notes.

\subsection{The D1-D5 system}

The first success in building this kind of correspondence between
black hole metrics and D-branes was achieved by Strominger and Vafa
\cite{SV}. They considered 5-dimensional
supergravity obtained by compactifying 10-dimensional type IIB theory
on a 5-dimensional compact manifold (for example, the 5-torus), 
and constructed a class of black holes carrying
2 separate $U(1)$ charges. These solutions may be viewed as generalizations
of the well-known 4-dimensional charged (Reissner-Nordstrom) black hole.
For the Reissner-Nordstrom black hole the mass
is bounded from below by a quantity proportional to the charge.
In general, when the mass saturates the lower (BPS)
bound for a given choice
of charges, then the black hole is called extremal.
The extremal Strominger-Vafa black hole
preserves $1/8$ of the supersymmetries present in vacuum.
Also, the black hole is constructed
in such a way that, just as for the Reissner-Nordstrom solution, 
the area of the
horizon is non-vanishing at extremality \ci{SV}. 
In general, an important quantity characterizing
black holes is the Bekenstein-Hawking entropy which is proportional
to the horizon area:
\be S_{BH} = {A_h\over 4G}
\ ,
\ee
where $G$ is the Newton constant.
Strominger and Vafa calculated the Bekenstein-Hawking entropy of their
extremal solution as a function of the charges and succeeded in reproducing
this result with D-brane methods. To build a D-brane system carrying the 
same set of charges as the black hole, they had to consider intersecting
D-branes wrapped over the compact 5-dimensional manifold. 
For example, one may consider D3-branes intersecting over a line or
D1-branes embedded inside D5-branes. The $1+1$ dimensional gauge theory
describing such an intersection is quite complicated, but the degeneracy
of the supersymmetric BPS states can nevertheless be calculated in the
D-brane description valid at weak coupling. 
For reasons that will become clear shortly,
the description in terms of black hole metrics is valid only at very
strong coupling. Luckily, due to the supersymmetry,
the number of states does not change as the coupling is increased.
This ability to extrapolate the D-brane counting to strong coupling
makes a comparison with the Bekenstein-Hawking entropy possible,
and exact agreement is found in the limit of large charges \cite{SV}.
In this sense the collection of D-branes provides a ``microscopic''
explanation of the black hole entropy.

This correspondence was quickly generalized to black 
holes slightly excited above the extremality \cite{cm,HSS}. 
Further, the Hawking radiation rates and the absorption cross-sections
were calculated and successfully reproduced by D-brane models
\cite{cm,wad}.
Since then this system has been
receiving a great deal of attention.
However, some detailed comparisons are hampered by the complexities of the
dynamics of intersecting D-branes: to date there is no 
first principles approach to the Lagrangian of the $1+1$ dimensional 
conformal field theory on the intersection. 
For this and other reasons it has turned out 
very fruitful to study a similar correspondence
for simpler systems which involve parallel D-branes 
only \cite{gkp,ENT,kleb,gukt,gkThree}. We turn to this subject
in the next section.

\subsection{Coincident D$p$-branes}

Our primary motivation is that, as explained above, parallel
D$p$-branes realize $p+1$ dimensional $U(N)$ SYM theories, and we may
learn something new about them from comparisons with Ramond-Ramond
charged black $p$-brane
classical solutions. These solutions in type II supergravity have been
known since the early 90's \cite{hs,DL}.
The metric and dilaton backgrounds may be expressed in the following
simple form:
\be
\label{metric}
   ds^2 =
H^{-1/2}(r)
    \left[ - f(r) dt^2 + \sum_{i=1}^p (d x^i)^2 \right] +
H^{1/2}(r)
    \left[f^{-1} (r) dr^2 + r^2 d\Omega_{8-p}^2 \right] \ ,
\ee
$$ e^\Phi = H^{(3-p)/4}(r)\ ,
$$
where
$$   H(r)  = 1 + {L^{7-p} \over r^{7-p}} \ , \qquad \  \  
f(r) = 1- {r_0^{7-p}\over
r^{7-p}}
\ ,
$$
and $d\Omega_{8-p}^2$ is the metric of a unit $8-p$ dimensional sphere.
The horizon is located at $r=r_0$ and the extremality
is achieved in the limit $r_0 \rightarrow 0$.
Just like the stacks of parallel D-branes, the extremal solutions are 
BPS saturated: they preserve 16 of the 32 supersymmetries present
in the type II theory. 
A solution with $r_0\ll L$ is called near-extremal.

The correspondence between the entropies of the
near-extremal $p$-brane solutions (\ref{metric}) and those
of the $p+1$ dimensional SYM theories was first considered in
\cite{gkp,ENT}.
In contrast to the 
situation encountered for the Strominger-Vafa black hole,
the Bekenstein-Hawking entropy vanishes in the extremal limit:
for $r\ll L$ the longitudinal part of the metric behaves as
$(r/L)^{(7-p)/2}\sum_{i=1}^p (d x^i)^2$ and hence the area of the horizon
vanishes.
The same is obviously true on the D-brane side because the stack of
D-branes is in its ground state. 
For $r_0>0$ the $p$-brane carries some excess
energy $E$ above its extremal value, and the
Bekenstein-Hawking entropy is also non-vanishing. 
The Hawking temperature is then defined by
$ T^{-1} = \partial S_{BH}/\partial E$.
A different, but equivalent, way of calculating the Hawking temperature
is to consider a Euclidean continuation of the solutions
(\ref{metric}). The Euclidean time takes values on a circle of
circumference $\beta = 1/T$. As shown by Gibbons and Hawking \cite{Gib},
$\beta$ has to be chosen in such a way that the geometry has no
conical singularity at the horizon. We will use this method below to
derive $T$ as a function of $r_0$ and $R$ in one step.

Among the solutions (\ref{metric})
$p=3$ has a special status: in the extremal limit $r_0 \rightarrow 0$
the 3-brane solution 
\be
\label{geom}
ds^2 = \left (1+{L^4\over r^4}\right )^{-1/2}
\left (- dt^2 +dx_1^2+ dx_2^2+ dx_3^2\right )
+ \left (1+{L^4\over r^4}\right )^{1/2}
\left ( dr^2 + r^2 d\Omega_5^2 \right )\ 
\ee
is perfectly non-singular \cite{gt}.
One piece of evidence is that the dilaton $\Phi$ is constant for
$p=3$ but blows up at $r=0$ for all other extremal solutions.
Furthermore, the limiting form of the extremal
metric as $r\rightarrow 0$ is
\be \label{adsmetric}
ds^2 = {L^2 \over z^2} \left( -dt^2 + d\vec{x}^2 + dz^2 \right) +
    L^2 d\Omega_5^2 \ ,
\ee
where $z={L^2\over r}$. This describes the direct product of
5-dimensional Anti-de Sitter space,
$AdS_5$, and the 5-dimensional sphere, ${\bf S}^5$,
with equal radii of curvature $L$ \cite{gt}. 
To be more precise, the above metric with $z$ ranging from
$0$ to $\infty$ does not cover the entire $AdS_5$ space, but only
its Poincar\' e wedge. This space has a horizon located at 
infinite $z$ ($r=0$). After a Euclidean continuation we obtain
the entire Euclidean $AdS_5$ space also known as the 
Lobachevsky space ${\bf L}_5$.

Since both factors of the $AdS_5\times {\bf S}^5$ space (\ref{adsmetric})
are maximally symmetric, we have
\be
R_{abcd}= -{1\over L^2} [g_{ac} g_{bd}- g_{ad} g_{bc}]
\ee 
for the $AdS_5$ directions, and
\be
R_{ijkl}= {1\over L^2} [g_{ik} g_{jl}- g_{il} g_{jk}]
\ee
for the ${\bf S}^5$ directions.
This shows that
near $r=0$ the extremal 3-brane geometry 
(\ref{geom}) is non-singular and, in fact, all 
appropriately measured curvature
components become small for large $L$.
Roughly speaking, this geometry may be 
viewed as a semi-infinite throat of radius $L$ which for
$r \gg L$ opens up into flat $9+1$ dimensional space.
Thus, for $L$ much larger than the string scale $\sqrt {\alpha'}$, 
the entire 3-brane geometry has small curvatures
everywhere and is appropriately described by the supergravity
approximation to type IIB string theory.

Let us see how the requirement $L\gg \sqrt{\alpha'}$ translates into
the language of $U(N)$ SYM theory on $N$ coincident D3-branes.
To this end it is convenient to equate the ADM tension of the
extremal 3-brane classical solution to $N$ times the tension
of a single D3-brane. In this fashion we find the relation
\cite{gkp}
\be \label{quantiz}
{2\over \kappa^2} L^4 \Omega_5 = N{\sqrt\pi\over \kappa}
\ ,
\ee
where $\Omega_5 = \pi^3$ is the volume of a unit 
5-sphere, and
$\kappa=\sqrt{8\pi G}$ is the 10-dimensional gravitational 
constant. It follows that
\be\label{throatrel}
L^4 = {\kappa\over 2\pi^{5/2}} N \ .
\ee 
Since $\kappa = 8 \pi^{7/2} g_{st} \alpha'^2$, 
(\ref{throatrel}) gives
$ L^4 =4\pi N g_{st} \alpha'^2$. In turn, $g_{st}$ determines
the Yang-Mills coupling on the D3-branes through 
$g_{\rm YM}^2 = 4\pi g_{st}$. Thus, we have
\be
L^4 = g_{\rm YM}^2 N \alpha'^2 ,\ee
i.e. the size of the throat in string units is measured by the
`t Hooft coupling! This remarkable emergence of the `t Hooft coupling
from gravitational considerations is at the heart of the success of
the AdS/CFT correspondence. Moreover,
the requirement $L\gg \sqrt{\alpha'}$ translates into
$g_{\rm YM}^2 N \gg 1$: the gravitational approach is valid when
the `t Hooft coupling is very strong and the traditional field
theoretic methods are not applicable. 

\subsection{Entropy of Near-extremal 3-branes}

Now consider the near-extremal 3-brane geometry. In the near-horizon region,
$r \ll L$, we may replace $H(r)$ by $L^4/r^4$. 
The resulting metric \ci{jthroat}
\begin{equation}
\label{throatmetric}
   ds^2 = {r^2 \over L^2} 
    \left[ -  \left (1- {r_0^4\over r^4} \right )
dt^2 + d\vec{x}^2 \right] +
    {L^2 \over r^2} 
  \left  ( 1- {r_0^4\over r^4} \right )^{-1}  dr^2 + L^2 d\Omega_5^2 \ ,
\end{equation}
is a product of ${\bf S}^5$  with  a  certain
limit of the Schwarzschild black hole in $AdS_5$
 \cite{newWit}. The Euclidean Schwarzschild black hole is asymptotic to
${\bf S}^1\times {\bf S}^3$, and the required limit is 
achieved as the volume of
${\bf S}^3$ is taken to infinity. Thus, the Euclidean continuation of
the metric (\ref{throatmetric}) is asymptotic to 
${\bf S}^1\times {\bf R}^3$. 
To determine the circumference
of ${\bf S}^1$, $\beta=1/T$, it is convenient 
to set $r= r_0(1  + L^{-2} \rho^2)$. For small $\rho$ 
the relevant 2d  part of the Euclidean  metric is:
\begin{equation}
ds^2 = d\rho^2 + {4r_0^2\ov L^4} \rho^2 d\tau^2\ , \ \  \qquad \tau=it \ .
\end{equation}
In order to avoid a conical singularity at the horizon,
the period of the Euclidean time
has to be $\beta = \pi L^2/r_0$.

The 8-dimensional ``area'' of the horizon can be read off from the
metric (\ref{throatmetric}). If the spatial volume of the D3-brane (i.e.
the volume of the $x_1, x_2, x_3$ coordinates) is taken to be $V_3$,
then we find
\be
A_h = (r_0/L)^3 V_3 L^5 \Omega_5 = \pi^6 L^8 T^3 V_3 \ .
\ee 
Using (\ref{throatrel}) we arrive at the Bekenstein-Hawking
entropy \cite{gkp}
\begin{equation}
\label{bhe}
S_{BH}= {2\pi A_h\over \kappa^2} = {\pi^2\over 2} N^2 V_3 T^3
\ .
\end{equation}
In \cite{gkp} this gravitational entropy of a near-extremal
3-brane of Hawking temperature $T$ was compared with the entropy
of the ${\cal N}=4$ supersymmetric $U(N)$ gauge theory (which lives
on $N$ coincident D3-branes) heated up to the same temperature. 
The results turned out to be quite interesting.

The entropy of a free $U(N)$ ${\cal N}=4$
supermultiplet, which consists of the gauge field, $6 N^2$ massless
scalars and $4 N^2$ Weyl fermions, can be calculated using
the standard statistical mechanics of a massless gas (the black body 
problem), and the answer is
\be \label{ffc}
S_0= {2 \pi^2\over 3} N^2 V_3 T^3
\ .
\ee
It is remarkable that the 3-brane geometry
captures the $T^3$ scaling characteristic of a conformal
field theory (in a CFT this scaling is guaranteed by the extensivity of
the entropy and the absence of dimensionful parameters).
Also, the $N^2$ scaling indicates the presence of $O(N^2)$
unconfined degrees of freedom, which is exactly what we expect in
the ${\cal N}=4$ supersymmetric $U(N)$ gauge theory.
On the other hand, the relative factor of $3/4$ between $S_{BH}$ and
$S_0$ at first appeared mysterious.
In fact, this factor
is not a contradiction but rather a prediction about the strongly
coupled ${\cal N}=4$ SYM theory at finite temperature. 

Indeed, as we argued above, the supergravity calculation of the
Bekenstein-Hawking entropy, (\ref{bhe}),
is relevant to the $g_{\rm YM}^2 N\rightarrow\infty$ limit of the 
${\cal N}=4$ $SU(N)$ gauge theory,
while the free field calculation, (\ref{ffc}), 
applies to the $g_{\rm YM}^2 N\rightarrow
0$ limit. Thus, the relative factor of $3/4$ is not a discrepancy:
it relates two different limits of the theory. 
Indeed, on general field theoretic
grounds we expect that in the `t Hooft large $N$
limit the entropy is given by \cite{GKT}
\be
S= {2 \pi^2\over 3} N^2 f(g_{\rm YM}^2 N) V_3 T^3
\ .\ee
The function $f$ is certainly not constant: for example, recent
calculations \cite{foto} show that its perturbative expansion is
\be \label{weak} 
f(g_{\rm YM}^2 N) = 1 - {3\over 2\pi^2} g_{\rm YM}^2 N 
+{3+\sqrt 2\over \pi^3} \left ( g_{\rm YM}^2 N \right )^{3/2} + \ldots
\ee
Thus, the Bekenstein-Hawking entropy in supergravity, (\ref{bhe}),
is translated into the prediction that 
\be
\lim_{g_{{\rm YM}}^2 N \rightarrow \infty}
f(g_{{\rm YM}}^2 N ) = {3\over 4}\ .
\ee 

Furthermore, string theoretic corrections to the supergravity action
may be used to develop a strong coupling expansion around this
limiting value. The first such correction comes from the
leading higher derivative term in the type IIB effective action:
\be
I=
 -{1\over 16\pi G} \int d^{10} x \sqrt g
\ \bigg[ R - \half (\del \phi)^2 - {1 \ov 4 \cdot 5!}   (F_5)^2  +...+ 
\  \g \ e^{- {3\ov 2} \phi}  W + ...\bigg]   \  ,
\la{aaa}
\ee
where
$$ \ \ \ \ \ \   
  \g= { 1\ov 8} \zeta(3)(\alpha')^3 \ , 
$$
and $W$ depends only on the Weyl tensor:
\be
W =  C^{hmnk} C_{pmnq} C_{h}^{\ rsp} C^{q}_{\ rsk} 
 + \half  C^{hkmn} C_{pqmn} C_h^{\ rsp} C^{q}_{\ rsk}  
 \  . 
\la{rrrr}
\ee
The value of the supergravity action
should be identified with the free energy of the thermal 
gauge theory \cite{newWit}.
The first correction to the free energy may be found by evaluating
$O(\alpha'^3)$ term on the leading order metric (\ref{throatmetric})
and this gives \cite{GKT}
\eqn{GotFT}{
   F = -{\pi^2 \over 8} N^2 V_3 T^4 \left (1 + 
   {15\over 8}(g_{{\rm YM}}^2 N)^{-3/2} \right ) \ .
  }
 Via the standard thermodynamics relation
$S= -{\partial F\over \partial T}$ this translates into the following 
form of the function $f$ for large `t Hooft coupling, 
\be \label{strong}
 f(g_{{\rm YM}}^2 N) = {3\over 4} +
{45\over 32} \zeta(3) (g_{{\rm YM}}^2 N)^{-3/2}  + \ldots
\ .
\ee
In \cite{GKT} it was conjectured that
$f(g_{{\rm YM}}^2 N)$ is actually a monotonically decreasing function
which interpolates between 1 at $g_{{\rm YM}}^2 N=0$ and
$3/4$ at $g_{{\rm YM}}^2 N=\infty$. The monotonicity is consistent both with
the weak coupling behavior (\ref{weak}) calculated perturbatively,
and with the strong coupling behavior (\ref{strong}) found
using the dual string theory.

\section{Thermodynamics of M-branes}

Other examples of the ``conformal'' behavior of the
Bekenstein-Hawking entropy include the 11-dimensional 5-brane and
membrane solutions \cite{ENT}.
The microscopic description of
the 5-brane solution is in terms of a large number $N$ of coincident singly
charged 5-branes of M-theory, whose chiral world volume theory has
$(0,2)$ supersymmetry. Similarly, the membrane solution 
describes the large $N$ behavior of the CFT on $N$ coincident
elementary membranes, which is related to the strong coupling
limit of maximally supersymmetric $SU(N)$ Yang-Mills theory in
$2+1$ dimensions. In this section we read off the thermodynamics
of these CFT's from dual supergravity using the same method as
applied to the ${\cal N}=4$ SYM theory in the preceding section.
One interesting difference in the final answer
is that in M-theory there is no analogue
of the string scale parameter
$\alpha'$: all scales are set by the Planck mass. Correspondingly,
the only parameter in the CFT's living on M-branes is $N$: there
is no marginal direction similar to changing the `t Hooft
coupling of SYM theory. Therefore, corrections coming from higher
powers of curvature in the effective action are suppressed by
powers of $N$.

Let us start with   the M5-brane case.
The throat  limit of black M5-brane metric  (see, e.g., \ci{DLP})
is 
\be ds^2= {r\over L} (- f dt^2 + \sum_{i=1}^5 dx_i^2)+
{L^2\over r^2} f^{-1}dr^2  + L^2 d\Omega_4^2
\ ,
\ee
where $f= 1- {r_0^3\ov r^3}$.
Introducing the variable $y= \sqrt{L r}$, 
we bring the metric into the form
used in \ci{newWit} to describe a black hole in $AdS_{p+2}$
(cf. \rf{throatmetric})
 \be
 ds^2= {y^2\over L^2} \bigg[- (1 - {y^6_0 \ov y^6} ) dt^2 +  \sum_{i=1}^5 dx_i^2\bigg]+
4 {L^2\over y^2} (1 - {y^6_0 \ov y^6} )^{-1}
  dy^2  + L^2 d\Omega_4^2
\ , \ \ \ \ \ \   y_0 \equiv  (L r_0)^{1/2} \ .  \la{yy}
\ee
As follows from the condition of regularity of the Euclidean metric,
the Hawking temperature is 
\be T = {3y_0\ov  4\pi L^2}
\ .
\ee
The entropy is calculated from the horizon area of this black hole,
and using the charge quantization for M5 branes \cite{ENT}, 
\be 
L^9= N^3 {\kappa_{11}^2\over 2^7 \pi^5}\ ,
\ee
we find that $S\sim N^3 V_5 T^5$. This suggests that
there are order $N^3$ degrees of freedom in this CFT, i.e.
their number grows with $N$ faster than in the Yang-Mills theory. 
In \cite{ENT} it was suggested that these degrees of freedom come from
small ``plumbing fixture'' M2-branes which can connect three M5-branes
at a time. The precise origin of the $N^3$ growth is not well-understood
and poses an interesting challenge.

The M-theory action contains an $R^4$ correction term, and its effect on
the free-energy of the theory on $N$ M5-branes 
turns out to be of order $N$ \cite{GKT}.\footnote{A similar
order $N$ correction to the $N^3$ result also appears in anomaly
calculations for the $(2,0)$ theory \cite{Harv,Tsey}.}
The conclusion is that the free energy of the large $N$
$(2,0)$ theory may be written as
\be
 F = -  V_5 T^6\ (a_0 N^3 + a_1 N+ \ldots)
\ , 
\ee
where
\be
a_0=2^6 3^{-7} \pi^3\ ,\qquad 
a_1  = \ 730\ ({2\pi\ov 3})^8\  ({\pi\ov 2})^{1/3}
\ .
\ee

In  the M2 brane case  the  throat  metric is
\be
 ds^2= {r^4\over L^4} (- f dt^2 + \sum_{i=1}^2 dx_i^2)+
{L^2\over r^2} f^{-1}dr^2  + L^2 d\Omega_7^2
\ , \label{MTwoThroat}
\ee
where $f= 1- {r_0^6\ov r^6}$.
The variable $y= {r^2\ov L}$ brings \eno{MTwoThroat} into the standard form
for a black hole in $AdS_4$:
\be
 ds^2= {y^2\over L^2} \bigg[ - (1- {y_0^3\ov y^3} ) 
 dt^2 + \sum_{i=1}^2 dx_i^2\bigg]+
{L^2\over 4 y^2} (1- {y_0^3\ov y^3} )^{-1}dy^2 
 + L^2 d\Omega_7^2
\  , \ \ \ \ \ \ \ \   y_0 \equiv {r_0^2 \ov L} \ . 
\ee
Calculating the Hawking temperature we find
\be  T=
 {3y_0\ov  2\pi L^2} \ . \ee
The entropy and free energy are calculated analogously to the
previous cases. Using the relation that follows from 
the M2 brane charge quantization \cite{ENT},
\be 
L^9= N^{3/2} {\kappa_{11}^2 \sqrt 2\over \pi^5}\ ,
\ee
we finally have
\be F = - V_2 T^3\  (b_0 N^{3/2} + b_1 N^{1/2})
\ , 
\ee
where
\be
b_0= 2^{7/2} 3^{-4} \pi^2\ ,
\qquad b_1 =\  64 \ ({2\pi \ov 3})^5 \   2^{1/6}\  \pi^{7/3}\ .
\ee
The $O(N^{3/2})$ term originates from the 2-derivative action, while
the $O(N^{1/2})$ term from the $R^4$ correction. Microscopic origins
of these scalings remain to be understood.

\section{From absorption cross-sections to two-point correlators}
 
A natural step beyond
the comparison of entropies is to interpret
absorption cross-sections for massless particles
in terms of the D-brane world volume theories \cite{kleb} 
(for 5-d black holes
the D-brane approach to absorption was initiated in \cite{cm,wad}).
For $N$ coincident D3-branes it is interesting
to inquire to what extent the supergravity and the weakly coupled
D-brane calculations agree. For example, they might scale differently
with $N$ or with the incident energy. Even if the scaling exponents 
agree, the overall normalizations
could differ by a subtle numerical factor similar
to the $3/4$ found for the 3-brane entropy.
Surprisingly, the low-energy absorption cross-sections
turn out to agree exactly \cite{kleb}.

To calculate the absorption cross-sections in the D-brane formalism one
needs the low-energy world volume action for coincident D-branes coupled to
the massless bulk fields. Luckily, these couplings may be deduced
from the D-brane Born-Infeld action. For example, the coupling
of 3-branes to the dilaton $\Phi$, the Ramond-Ramond scalar $C$,
and the graviton $h_{\alpha\beta}$ is given by \cite{kleb,gukt}
\be \label{sint}
   S_{\rm int} = {\sqrt \pi\over\kappa}
\int d^4 x \, \bigg[ \tr \left(
 \tf{1}{4}   {\Phi} F_{\alpha\beta}^2  -
  \tf{1}{4}   {C} F_{\alpha\beta} \tilde {F}^{\alpha\beta} \right)
+ \tf{1}{2} h^{\alpha\beta} T_{\alpha\beta} \bigg] \ ,
 \ee
 where $T_{\alpha\beta}$ is
the stress-energy tensor of the ${\cal N}=4$ SYM theory.
Consider, for instance, absorption of a dilaton incident on the 3-brane
at right angles with a low energy
$\omega$. Since the dilaton couples to 
${1\over 4 g_{\rm YM}^2}\tr F_{\alpha\beta}^2$
it can be converted into a pair of back-to-back gluons on the world volume.
The leading order calculation of the cross-section
for weak coupling gives \cite{kleb}
\be\label{absorb}
   \sigma = {\kappa^2  \omega^3 N^2\over 32 \pi} \ .
 \ee
The factor $N^2$ comes from the degeneracy of
the final states which is the number of different gluon species.

This result was compared with the absorption cross-section by
the extremal 3-brane geometry (\ref{geom}).
As discussed above, the geometry
is a non-singular semi-infinite throat which opens up at large
$r$ into flat 10-dimensional space. 
Waves incident from the $r \gg L$ region partly reflect back and
partly penetrate into the the throat region $r \ll L$.
The relevant s-wave radial equation turns out to be \cite{kleb}
\be
\label{Coulthree}
\left [{d^2\over d \rho^2} - {15\over 4 \rho^2}
+1 + {(\omega L)^4\over \rho^4} \right ] \psi(\rho) =0\ ,
\ee
where $\rho = \omega r$. For a low energy $\omega \ll 1/L$ we find
a high barrier separating the two asymptotic regions.
The low-energy behavior of the tunneling probability may be calculated
by the so-called matching method, and the resulting absorption
cross-section is \cite{kleb}
\be
\label{three}
\sigma_{SUGRA}= {\pi^4\over 8}\omega^3 L^8 \ .
\ee
Substituting (\ref{throatrel}) we find that the supergravity
absorption cross-section agrees exactly with the D-brane one,
without any relative factor like $3/4$.

This surprising result needs an
explanation. The most important question is: what is the range of
validity of the two calculations? 
The supergravity approach may be trusted only if
the length scale of the 3-brane solution is much larger than the 
string scale $\sqrt{\alpha'}$. As we have shown, this translates into
$N g_{st} \gg 1$.
Of course, the incident energy also has
to be small compared to $1/\sqrt{\alpha'}$. 
Thus, the supergravity calculation should
be valid in the ``double-scaling limit'' \cite{kleb}
\begin{equation}
\label{dsl}
{L^4\over \alpha'^2} = 4\pi  g_{st} N \rightarrow \infty\ ,
\qquad\qquad \omega^2 \alpha' \rightarrow 0\ .
\end{equation}
If the description of the black 3-brane by a stack of
many coincident D3-branes is correct, 
then it {\it must} agree with the supergravity results in this limit,
which corresponds to {\it infinite} `t Hooft coupling in the  
${\cal N}=4$ $U(N)$ SYM theory. Since we also want to send $g_{st}
\rightarrow 0$ in order to suppress the string loop corrections,
we necessarily have to take the large $N$ limit.

Although we have sharpened the region of applicability of the
supergravity calculation (\ref{three}), we have not yet explained why
it agrees with the leading order perturbative result (\ref{absorb})
on the D3-brane world volume. 
After including the higher-order SYM corrections, the general structure
of the absorption cross-section in the large $N$ limit
is expected to be \cite{gkThree}
\be\label{newabsorb}
   \sigma = {\kappa^2  \omega^3 N^2\over 32 \pi} a(g_{\rm YM}^2 N)\ ,
 \ee
where
$$ a(g_{\rm YM}^2 N) = 1 + b_1 g_{\rm YM}^2 N + 
b_2 (g_{\rm YM}^2 N)^2 + \ldots
$$
For agreement with supergravity, the strong
`t Hooft coupling limit of $a(g_{\rm YM}^2 N)$ should be equal to 1 
\cite{gkThree}. In fact, a stronger result is true: all perturbative
corrections vanish and $a=1$ independent of the coupling.
This was first shown explicitly in \cite{gkThree} for the graviton
absorption. The absorption cross-section for a graviton polarized along
the brane, say $h_{xy}$, is related to the discontinuity across the
real axis (i.e. the absorptive part) of the two-point function
$\langle T_{xy} (p) T_{xy} (-p) \rangle$
in the SYM theory. In turn,
this is determined by a conformal ``central charge''
which satisfies a non-renormalization theorem: it is completely
independent of the `t Hooft coupling.

In general, the two-point function of a gauge invariant operator in
the strongly coupled SYM theory may be read off from the
absorption cross-section for the supergravity field which
couples to this operator in the world volume action \cite{gkThree,KTV}.
Consider, for instance, scalar operators.
For a canonically normalized bulk scalar field coupling to the
brane through an interaction
\be
S_{\rm int} = \int d^4 x \phi (x,0) {\cal O} (x)
\ee
($\phi(x,0)$ denotes the value of the field at the transverse
coordinates where the D3-branes are located)
the precise relation is given by
\be
\label{eq:disc}
\sigma = \left. {1 \over 2 i \omega}  {\rm Disc}\; \Pi (p) \right|_{-p^2
=
\omega^2 - i \epsilon}^{-p^2 = \omega^2 + i \epsilon}
\ee
Here $\omega$ is the energy of the particle, and
\be
\Pi(p) = \int d^4 x e^{i p \cdot x} \langle {\cal O}(x) {\cal O} (0)
\rangle
\ee
which depends only on $s=-p^2$. To evaluate (\ref{eq:disc}) we extend
$\Pi$ to  complex values of $s$ and compute the discontinuity of $\Pi$
across the real axis at $s=\omega^2$.

Some examples of the field operator correspondence may
be read off from (\ref{sint}). Thus, we learn that the dilaton
absorption cross-section measures the normalized 2-point function
$\langle O_\Phi  (p) O_\Phi (-p) \rangle$ where $O_\Phi$
is the operator that couples to the dilaton: 
\be
O_\Phi = {1\over 4 g_{\rm YM}^2 }\tr F_{\alpha\beta} F^{\alpha\beta} + \ldots 
\ee
(we have not written out the scalar and fermion terms explicitly).
Similarly,
the Ramond-Ramond scalar absorption cross-section measures  
$\langle O_C  (p) O_C (-p) \rangle$ where 
\be
O_C = 
{1\over 4 g_{\rm YM}^2 }\tr F_{\alpha\beta} \tilde {F}^{\alpha\beta} 
+\ldots \ee
The agreement of these two-point functions
with the weak-coupling calculations performed in
\cite{kleb,gukt} is explained by non-renormalization
theorems related by supersymmetry to the
non-renormalization of the central charge discussed in
\cite{gkThree}. 
Thus, the proposition that the $g_{\rm YM}^2 N\rightarrow
\infty$ limit of the large $N$ ${\cal N}=4$ SYM theory can be extracted from 
the 3-brane of type IIB supergravity has passed its first consistency checks.

It is of further interest to perform similar comparisons in cases
where the relevant non-renormalization theorems have not yet been
proven. Consider, for instance, absorption of the dilaton in the
$l$-th partial wave. Now the angular Laplacian on ${\bf S}^5$ has the
eigenvalue $l(l+4)$ and the effective radial equation becomes
\be
\label{partialthree}
\left [{d^2\over d \rho^2} - {l(l+4) + 15/4\over \rho^2}
+1 + {(\omega L)^4\over \rho^4} \right ] \psi(\rho) =0\ ,
\ee
The thickness of the barrier through which the particle has to
tunnel increases with $l$, and we expect the cross-section
to become increasingly suppressed at low energies.
Indeed, a detailed matching calculation \cite{gukt,KTV} gives  
\begin{equation}
\sigma^l_{SUGRA} = \frac{\pi^4}{ 24}  \frac{(l + 3) (l + 1)}{ [(l + 1)!]^4}
\left( \frac{\omega L}{ 2} \right)^{4l} \omega^3 L^8.
\end{equation}
After 
replacing $L^4$ through (\ref{throatrel}) this can be rewritten as
\begin{equation}
\sigma^l = \frac{N^{l + 2} \kappa^{l + 2} \omega^{4l + 3} (l + 3)}{
3 \cdot 2^{5l + 5} \pi^{5l/2 + 1}l![(l + 1)!]^3}.
\label{eq:semiclassical-absorption}
\end{equation}
What are the operators whose 2-point functions are related to these
cross-sections? For a single D3-brane one may expand the
dilaton coupling in a Taylor series in the transverse coordinates to
obtain the following bosonic term \cite{kleb}:
\be
{1\over 4 l!} F_{\alpha\beta} F^{\alpha\beta} X^{(i_1}\ldots X^{i_l)}
\ ,
\ee
where the parenthesis pick out a transverse traceless polynomial,
which is an irreducible representation of $SO(6)$.
The correct non-abelian generalization of this term is
\cite{KTV}
\be \label{incomplete}
{1\over 4 l!} {\rm STr} 
\left [F_{\alpha\beta} F^{\alpha\beta} X^{(i_1}\ldots X^{i_l)}
\right ]
\ ,
\ee
where STr denotes a symmetrized trace \cite{Tseytlin}: 
in this particular case we have to average
over all positions of the $F$'s modulo cyclic permutations.
A detailed calculation in \cite{KTV} reveals that the 2-point function
of this operator calculated at weak coupling
accounts for ${6\over (l+2)(l+3)} $ 
of the semiclassical absorption cross-section
(\ref{eq:semiclassical-absorption}) in the sense of the relation
(\ref{eq:disc}). Luckily, (\ref{incomplete}) is not the complete
world volume coupling to the dilaton in the $l$-th partial wave.
${\cal N}=4$ supersymmetry guarantees that there are additional terms
quadratic and quartic in the fermion fields. When all these terms are
taken into account there is {\it exact} agreement between the weak
and strong coupling calculations of the 2-point functions.
This strongly suggests that the complete $l$-th
partial wave operators are protected by supersymmetric
non-renormalization theorems. Proving them is an interesting challenge
(for recent progress, see \cite{Sken}).

\section{The AdS/CFT Correspondence}

The circle of ideas reviewed in the previous sections received
an important development by Maldacena \cite{jthroat} who also connected
it for the first time with the QCD string idea. Maldacena made
a simple and powerful observation that the ``universal'' region of
the 3-brane geometry, which should be directly identified with the
${\cal N}=4$ SYM theory, is the throat, i.e. the region $r\ll L$.\footnote{
Related ideas were also pursued in \cite{HYU}.} The
limiting form of the 3-brane metric (\ref{geom}) is (\ref{adsmetric})
which describes the space
$AdS_5\times {\bf S}^5$ with equal radii of curvature $L$.
One also finds that the self-dual 5-form R-R field strength
has $N$ units of flux through this space (the field strength term in the
Einstein equation effectively gives a positive cosmological 
constant on ${\bf S}^5$
and a negative one on $AdS_5$).
Thus, Maldacena conjectured that type IIB string theory on 
$AdS_5\times {\bf S}^5$
should be somehow dual to the 
${\cal N}=4$ supersymmetric
$SU(N)$ gauge theory.

By the same token, identifying, as we did in section 3,
the 2-brane and 5-brane
classical solutions
of 11-dimensional supergravity with stacks of M2 and M5 branes respectively
leads to similar dualities in the M-theory
context. In particular, a large $N$ 6-dimensional $(2,0)$ theory
is conjectured to be dual to the $AdS_7\times {\bf S}^4$ background,
and a large $N$ maximally supersymmetric 3-dimensional
gauge theory is conjectured to be dual to the $AdS_4\times {\bf S}^7$
background. In the following we will discuss only the
D3-brane case, but generalization to the M-branes is straightforward.

Maldacena's argument was based on the fact that the low-energy
($\alpha'\rightarrow 0$) limit may be taken directly in the 3-brane
geometry and is equivalent to the throat ($r \rightarrow 0$) limit.
Another way to motivate the identification of the gauge theory
with the throat is to think about the absorption of massless particles
considered in the previous section. In the D-brane description,
a particle incident from the asymptotic infinity 
is converted into an excitation of the stack of D-branes, i.e. into
an excitation of the gauge theory on the world volume.
In the supergravity description, a particle incident from the asymptotic 
(large $r$) region tunnels into the $r\ll L$ region and produces an excitation
of the throat. The fact that the two different descriptions of
the absorption process give identical cross-sections supports the 
identification of excitations of $AdS_5\times {\bf S}^5$ with the excited
states of the ${\cal N}=4$ SYM theory.

Another strong piece of support for this identification comes from
symmetry considerations \cite{jthroat}. The isometry group of
$AdS_5$ is $SO(2,4)$, and this is also the conformal group in
$3+1$ dimensions. In addition we have the isometries of ${\bf S}^5$ which
form $SU(4)\sim SO(6)$. This group is identical to the R-symmetry of
the ${\cal N}=4$ SYM theory. After including the fermionic generators
required by supersymmetry, the full isometry supergroup of the
$AdS_5\times {\bf S}^5$ background is $SU(2,2|4)$, which is identical to
the ${\cal N}=4$ superconformal symmetry.
We will see that in theories with reduced supersymmetry
the ${\bf S}^5$ factor becomes replaced by other
compact Einstein spaces $X_5$,
but $AdS_5$ is the ``universal'' factor present in the dual
description of any large $N$ CFT and realizing the $SO(2,4)$ conformal
symmetry. One may think of these backgrounds as type IIB theory
compactified on $X_5$ down to 5 dimensions. Such Kaluza-Klein
compactifications of type IIB supergravity were extensively studied in the
mid-eighties \cite{GRW,Romans,Duff}, and special attention was devoted to
the $AdS_5\times {\bf S}^5$ solution because it is
a maximally supersymmetric background \cite{SH,Kim}.
It is remarkable that these early works on compactification
of type IIB theory
were actually solving large $N$ gauge theories without knowing it.

As Maldacena has emphasized, it is also important to go beyond
the supergravity limit and think of the $AdS_5\times X_5$ space
as a background of string theory \cite{jthroat}. Indeed, type IIB strings
are dual to the electric flux lines in the gauge theory, and
this provides a natural set-up for calculating correlation 
functions of the Wilson loops \cite{Malda}. 
Furthermore, if $N$ is sent to infinity
while $g_{\rm YM}^2 N$ is held fixed and finite, then there are finite
string scale corrections to the supergravity limit \cite{jthroat,US,EW}
which proceed in powers of
${\alpha'\over L^2} = \left (g_{\rm YM}^2 N \right)^{-1/2}
.
$
If we wish to study finite $N$, then there are also string loop
corrections in powers of
$ {\kappa^2\over L^8} \sim N^{-2}
.
$
As expected, taking $N$ to infinity enables us to take
the classical limit of the string theory on $AdS_5\times X_5$.
However, in order to understand the large $N$
gauge theory with finite `t Hooft coupling,
we should think of $AdS_5\times X_5$ as the target space of a 
2-dimensional sigma model describing the classical string physics 
\cite{US}. The fact that after the compactification on $X_5$ 
the string theory is 5-dimensional supports Polyakov's idea \cite{Sasha}.
In $AdS_5$ the fifth dimension is related to
the radial coordinate and, after a change of variables
$z= L e^{-\varphi/L}$, the sigma model action turns into a special
case of the general ansatz proposed in \cite{Sasha},
\be
S = {1\over 2}\int d^2 \sigma [(\partial_\alpha \varphi)^2 + a^2 (\varphi)
(\partial_\alpha X^i)^2 + \ldots ]
\ ,
\ee
with $a(\varphi) = e^{\varphi/L}$. More generally, the ``warp factor''
$a^2(\varphi)$ describes RG flow of string tension in the
gauge theory \cite{Sasha}.
It is clear, however, that the string sigma models dual to the gauge
theories are of rather peculiar nature. The new feature revealed
by the D-brane approach, which is also a major stumbling block,
is the presence of the Ramond-Ramond background fields. Little is known
to date about such 2-dimensional field theories and, in spite of
recent new insights \cite{MT,BVW,Berk}, 
an explicit solution is not yet available.

\subsection{Correlation functions and the bulk/boundary correspondence}

Maldacena's work provided a crucial insight that the $AdS_5\times {\bf S}^5$ 
throat is the part of the 3-brane geometry that is most directly related
to the ${\cal N}=4$ SYM theory. It is important to go further, however,
and explain precisely in 
what sense the two should be identified and how physical information
can be extracted from this duality. Major strides towards
answering these questions were made in two subsequent papers 
\cite{US,EW} where
essentially identical methods for calculating correlation functions
of various operators in the gauge theory were proposed.
As we mentioned in section 2.2, even prior to \cite{jthroat} some
information about the field/operator correspondence
and about the two-point functions had been extracted from
the absorption cross-sections. The reasoning of \cite{US}
was a natural extension of these ideas. 

One may motivate the general method
as follows. When a wave is absorbed, it tunnels from the asymptotic
infinity into the throat region, and then continues to propagate
toward smaller $r$. Let us separate the 3-brane geometry into two
regions: $r\gsim L$ and $r\lsim L$. For $r\lsim L$ the metric is
approximately that of $AdS_5\times {\bf S}^5$, while for $r\gsim L$
it becomes very different and eventually approaches the flat metric.
Signals coming in from large $r$ (small $z=L^2/r$)
may be thought of as disturbing the
``boundary'' of $AdS_5$ at $r\sim L$, and then propagating into the
bulk. This suggests that, if we discard the $r\gsim L$ part of the
3-brane metric, then 
the gauge theory correlation functions are related to
the response of the string theory to boundary conditions at $r\sim L$.
Guided by this idea, \cite{US} proposed
to identify the generating functional of connected
correlation functions in the gauge theory with the extremum of the
classical string theory action $I$ subject to the boundary conditions 
that $\phi(x^\lambda, z) = \phi_0 (x^\lambda)$ at
$z=L$ (at $z=\infty$ all fluctuations are required to vanish):\foot{
As usual, in calculating correlation functions in a CFT
it is convenient to carry out the Euclidean continuation. On the string
theory side we then have to use the Euclidean version of $AdS_5$.} 
\be \label{GKPW}
   W[\phi_0 (x^\lambda)] = I_{\phi_0 (x^\lambda)} 
     \ . 
\ee 
 $W$ generates the connected Green's functions of the gauge theory
operator that corresponds to the field $\phi$ in the sense explained
in section 2.2, 
while $I_{\phi_0 (x^\lambda)} $ 
is the extremum of the classical string action
subject to the boundary conditions. 
An essentially identical prescription
was also proposed in \cite{EW} with a somewhat different motivation.
If we are interested in the
correlation functions at infinite `t Hooft coupling, then the
problem of extremizing the
classical string action reduces to solving the equations of
motion in type IIB supergravity whose form is known explicitly
\cite{SH}.

Our reasoning suggests
that from the point of view of the metric (\ref{adsmetric})
the boundary conditions are imposed not at $z=0$ 
(which would be a true boundary 
of $AdS_5$) but at some finite value $z= z_{cutoff}$.
It does not matter which value it is since 
the metric (\ref{adsmetric}) is unchanged by an
overall rescaling of the coordinates $(z, x^{\lambda})$; thus,
such a rescaling can take $z=L$ into $z= z_{cutoff}$ for any
$z_{cutoff}$.
The physical meaning of this cut-off is that it acts as a UV
regulator in the gauge theory \cite{US,sw}. Indeed, the radial coordinate of
$AdS_5$ is to be thought of as the effective energy scale of the gauge
theory \cite{jthroat}, and decreasing $z$ corresponds to increasing
the energy. 
A safe method for performing calculations of correlation
functions, therefore, is
to keep the cut-off on the $z$-coordinate at intermediate stages
and remove it only at the end \cite{US,freed}. This way the calculations are
not manifestly AdS-invariant, however. Usually there is another way to
regularize the action which is manifestly AdS invariant. Luckily, when
all subtleties are taken into account, these two ways of performing 
calculations do agree \cite{KWnew,Haro}.

\subsection{Two-point functions}

Below we present a brief discussion of two-point functions
of scalar operators. The corresponding field in $AdS_{d+1}$ is a scalar
field of mass $m$ with the action
\be
{1\over 2} \int d^{d+1} x \sqrt g
\left [ g^{\mu\nu} \partial_\mu \phi \partial_\nu \phi
+ m^2 \phi^2 \right ]=
{1\over 2} \int d^d x dz  z^{-d+1}\left [(\partial_z \phi)^2+
(\partial_i\phi)^2  + {m^2\over z^2} \phi^2 \right ]
\ ,
\ee
where we have set $L=1$.
In calculating correlation functions of vertex
operators from the AdS/CFT correspondence,
the first problem is to reconstruct an on-shell field in $AdS_{d+1}$
from its boundary behavior. 
The small $z$ behavior of the classical solution is
\be \label{bc}
\phi (z, \vec x)\rightarrow z^{d-\Delta}
[\phi_0 (\vec x) + O(z^2)]
+z^\Delta [A(\vec x) + O(z^2)] \ ,
\ee
where $\Delta$ is one of the roots of
\be \label{relation}
\Delta (\Delta -d) = m^2\ .
\ee
$\phi_0 (\vec x)$ is regarded as a ``source'' function and
$A(\vec x)$
describes a physical fluctuation.

It is possible to regularize the Euclidean
action \cite{KWnew}
to obtain the following value as a functional of the source,
\be
I(\phi_0) =- (\Delta- (d/2)) \pi^{-d/2}
{\Gamma (\Delta)\over \Gamma (\Delta-
(d/2))}
\int d^d \vec x \int d^d \vec x'
{\phi_0 (\vec x) \phi_0 (\vec x')\over |\vec x- \vec x'|^{2\Delta} }
\ .
\ee
Varying twice with respect to $\phi_0$ we find that the two-point
function of the corresponding operator is
\be \label{twopoint}
\langle {\cal O}(\vec x){\cal  O}(\vec x')\rangle=
{(2\Delta - d) \Gamma (\Delta)\over \pi^{d/2} \Gamma (\Delta -
(d/2))} {1\over
|\vec x- \vec x'|^{2\Delta} }
\ .
\ee
Precisely the same normalization of the two-point function follows from a
different regularization where $z_{cutoff}$ is kept at intermediate stages
\cite{US,freed}.

We note that $\Delta$ is the dimension of the operator.
Which of the two roots of (\ref{relation}) should we choose?
Superficially it seems that we should always choose
the bigger root,
\be
\label{dimen}
\Delta_+ = {d\over 2}+\sqrt{ {d^2\over 4} + m^2 }\ ,
\ee
because then the $\phi_0$ term in (\ref{bc}) dominates over the
$A$ term. For positive $m^2$, $\Delta_+$ is certainly the right
choice: here the other root, 
\be
\label{otherdimen}
\Delta_- = {d\over 2}-\sqrt{ {d^2\over 4} + m^2 }\ ,
\ee
is negative. However, it turns out
that for
\be \label{range}
-{d^2\over 4} < m^2 < - {d^2\over 4} + 1
\ee
both roots of (\ref{relation}) may be chosen. Thus, there are {\it two}
possible CFT's corresponding to the same classical AdS action
\cite{KWnew}:
in one of them the corresponding operator has dimension $\Delta_+$
while in the other -- dimension $\Delta_-$. (The fact that there are
two admissible boundary conditions in $AdS_{d+1}$ for a scalar field
with $m^2$ in the range (\ref{range}) has been known since the old work of
Breitenlohner and Freedman \cite{BF}.) 
This conclusion resolves the following
puzzle. $\Delta_+$ is bounded from below by $d/2$ but there is
no corresponding bound in $d$-dimensional CFT (in fact, as we will see,
there are examples of field theories with operators that violate this
bound). However, in the range (\ref{range}) $\Delta_-$ is bounded
from below by $(d-2)/2$, and this is precisely the unitarity bound
on dimensions of scalar operators
in $d$-dimensional field theory! Thus, the ability to have dimension
$\Delta_-$ is crucial for consistency of the AdS/CFT duality.

A question remains, however, as to what is the correct definition
of correlation functions in the theory with dimension $\Delta_-$.
The answer to this question is related to the physical interpretation
of the function $A(\vec x)$ entering the boundary behavior of the
field (\ref{bc}). As suggested in \cite{BKL} this function is related
to the {\it expectation value} of the operator ${\cal O}$.
The precise relation, which holds even after interactions are taken into
account, is \cite{KWnew}
\be
A(\vec x) = {1\over 2\Delta -d} \langle {\cal O} (\vec x) \rangle
\ .
\ee
Thus, from the point of view of the $d$-dimensional
CFT, $(2\Delta - d) A(\vec x)$
is the variable conjugate to $\phi_0(\vec x)$.
In order to interchange $\Delta$ and $d-\Delta$, it is clear from
(\ref{bc}) that we have to interchange $\phi_0$ and $(2\Delta - d) A$. 
This is a canonical transformation which for tree-level correlators
reduces to a Legendre transform. Thus, the generating functional of
correlators in the $\Delta_-$ theory may be obtained by Legendre
transforming the generating functional of correlators in the $\Delta_+$
theory \cite{KWnew}. This gives a simple and explicit prescription for defining
correlation functions of operators with dimension
$\Delta_-$. For the 2-point function, for example, we find 
that the formula (\ref{twopoint})
is correct for both definitions of the theory, i.e. it makes sense
for all dimensions above the unitarity bound,
\be
\Delta > {d\over 2} -1
\ .
\ee
Indeed, note that for such dimensions 
the two-point function (\ref{twopoint}) is positive,
but as soon as $\Delta$ crosses the unitarity bound,
(\ref{twopoint})
becomes negative signaling a non-unitarity of the theory.
Thus, appropriate treatment of fields in $AdS_{d+1}$ gives information
on 2-point functions completely consistent with expectations from 
CFT$_d$. The fact that the Legendre transform prescription of
\cite{KWnew} works properly for higher-point correlation functions
was recently demonstrated in \cite{MW}.

Whether string theory on $AdS_5 \times X_5$ contains fields with 
mass-squared
in the range (\ref{range}) depends on $X_5$. The example discussed
in section 6, $X_5=T^{1,1}$, turns out to contain such fields, and the
possibility of having dimension $\Delta_-$, (\ref{otherdimen}), 
is crucial for 
consistency of the AdS/CFT duality in that case. However, 
for $X_5={\bf S}^5$ which is dual
to the ${\cal N}=4$ large $N$ SYM theory, there are no such fields
and all scalar dimensions are given by (\ref{dimen}).

The operators in the ${\cal N}=4$ large $N$ SYM theory
naturally break up into two classes: those that correspond to the
Kaluza-Klein states of supergravity and those that correspond to 
massive string states.
To reinstate
$L$ we simply replace $m$ by $mL$ in the formula for
operator dimension, (\ref{dimen}). Since the radius of the ${\bf S}^5$
is $L$, the masses of the Kaluza-Klein states are
proportional to $1/L$. Thus,
the dimensions of the corresponding operators are independent of $L$ and
therefore independent of $g_{\rm YM}^2 N$.
On the gauge theory side this
is explained by the fact that the supersymmetry protects the dimensions
of certain operators from being renormalized: they are completely determined
by the representation under the superconformal symmetry 
\cite{hw,Ferrara}.
All families of the Kaluza-Klein states, which correspond to such
BPS protected operators, were classified long ago \cite{Kim}.
Correlation functions of such operators in the strong `t Hooft coupling
limit may be obtained from the dependence of the supergravity action
on the boundary values of corresponding Kaluza-Klein fields, as in
(\ref{GKPW}). A variety of explicit calculations have been performed for
2-, 3- and even 4-point functions. The 4-point functions are
particularly interesting because their dependence on operator positions
is not determined by the conformal invariance. For state-of-the-art
results on them, see \cite{DFM,Arut}.

On the other hand, the masses of string excitations are
$m^2 = {4 n\over \alpha'}$ where $n$ is an integer.
For the corresponding operators the formula (\ref{dimen})
predicts that the dimensions do depend on the `t Hooft coupling and,
in fact, blow up for large $g_{\rm YM}^2 N$ as
$2\left (n g_{\rm YM} \sqrt {N} \right )^{1/2}$ \cite{US}.
This is a highly non-trivial prediction of the AdS/CFT duality which
has not yet been verified on the gauge theory side.

\section{Conformal field theories and Einstein manifolds}

As we mentioned above, the duality between 
type IIB strings on $AdS_5\times {\bf S}^5$
and the ${\cal N}=4$ SYM is naturally generalized to dualities between
backgrounds of the form  $AdS_5\times X_5$ and other conformal gauge theories.
The 5-dimensional compact space $X_5$ is required to be a
positively curved Einstein manifold, i.e. one for which
$R_{\mu\nu}= \Lambda g_{\mu\nu}$ with $\Lambda>0$.
The number of supersymmetries in the dual gauge theory is determined
by the number of Killing spinors on $X_5$.

The simplest examples of $X_5$ are the orbifolds ${\bf S}^5/\Gamma$ where 
$\Gamma$ is a discrete subgroup of $SO(6)$ \cite{ks,lnv}. In these cases
$X_5$ has the local geometry of a 5-sphere.
The dual gauge theory is the IR
limit of the world volume theory on a stack of $N$ D3-branes placed at
the orbifold singularity of ${\bf R}^6/\Gamma$. Such theories
typically involve product gauge groups $SU(N)^k$ coupled to matter
in bifundamental representations \cite{dm}.

Constructions of the dual gauge theories for Einstein manifolds
$X_5$ which are not locally equivalent to ${\bf S}^5$ are also possible. 
The simplest
example is the Romans compactification on
$X_5= T^{1,1}= (SU(2)\times SU(2))/U(1)$ \cite{Romans,KW}. 
It turns out that
the dual gauge theory is the conformal limit of 
the world volume theory on a stack of $N$ D3-branes placed at
the singularity of a certain Calabi-Yau
manifold known as the conifold \cite{KW}. Let us explain this connection
in more detail. 

\subsection{D3-branes on the Conifold}

The conifold may be described by the
following equation in four complex variables,
\be \label{coni}
\sum_{a=1}^4 z_a^2 = 0
\ .
\ee
Since this equation is symmetric under an overall rescaling of the
coordinates, this space is a cone. Remarkably, the base of this cone
is precisely the space $T^{1,1}$ \cite{cd,KW}. A simple argument
in favor of this is based on the symmetries. In order to find the base
we intersect (\ref{coni}) with $\sum_{a=1}^4 |z_a|^2 = 1$.
The resulting space has the $SO(4)$ symmetry which rotates the $z$'s,
and also the $U(1)$ R-symmetry under $z_a\rightarrow e^{i\theta} z_a$.
Since $SO(4)\sim SU(2)\times SU(2)$ these symmetries coincide with those
of $T^{1,1}$. In fact, the metric on the conifold may be
cast in the form \cite{cd}
\be
ds_6^2 = dr^2 + r^2 ds_5^2\ ,
\ee
where 
\begin{equation} \label{co}
d s_5^2=
{1\over 9} \bigg(d\psi + 
\sum_{i=1}^2 \cos \theta_i d\phi_i\bigg)^2+
{1\over 6} \sum_{i=1}^2 \left(
d\theta_i^2 + {\rm sin}^2\theta_i d\phi_i^2
 \right)
\ .
\end{equation}
is the metric on $T^{1,1}$. Here $\psi$ is an angular coordinate
which  ranges from $0$ to $4\pi$,  while $(\theta_1,\phi_1)$
and $(\theta_2,\phi_2)$ parametrize two ${\bf S}^2$'s in a standard way.
Therefore, this form of the metric shows that
$T^{1,1}$ is an ${\bf S}^1$ bundle over ${\bf S}^2 \times {\bf S}^2$.

Now placing $N$ D3-branes at the apex of the cone we find the metric
\be
\label{newgeom}
ds^2 = \left (1+{L^4\over r^4}\right )^{-1/2}
\left (- dt^2 +dx_1^2+ dx_2^2+ dx_3^2\right )
+ \left (1+{L^4\over r^4}\right )^{1/2}
\left ( dr^2 + r^2 ds_5^2 \right )
\ee
whose near-horizon limit is $AdS_5\times T^{1,1}$ (once again,
$L^4\sim g_s N$). Thus, type IIB string theory on this space should
be dual to the infrared limit of the field theory on $N$ D3-branes
placed at the singularity of the conifold. Since Calabi-Yau spaces
preserve 1/4 of the original supersymmetries we find that this 
should be an ${\cal N}=1$
superconformal field theory. 
This field theory was constructed
in \cite{KW}: it is $SU(N)\times SU(N)$ gauge theory
coupled to two chiral superfields, $A_i$, in the $({\bf N}, \overline{\bf N})$
representation
and two chiral superfields, $B_j$, in the $(\overline{\bf N}, {\bf N})$
representation \cite{KW}. The $A$'s transform as a doublet under one
of the global $SU(2)$'s while the $B$'s transform
as a doublet under the other $SU(2)$.

A simple way to motivate the appearance of the fields $A_i,\ B_j$ is to
rewrite the defining equation of the conifold, (\ref{coni}), as
\begin{equation} \label{conifold}
\det_{i,j} z_{ij} = 0\ , 
\qquad z_{ij} ={1\over \sqrt{2}}\sum_n \sigma^n_{ij} z_n
\end{equation} 
where $\sigma^n$ are the Pauli matrices for $n=1,2,3$ 
and $\sigma^4$ is $i$ times the unit matrix.
This quadratic constraint may be ``solved'' by the substitution
\begin{equation}
z_{ij} = A_i B_j
\ ,
\end{equation}
where $A_i,\ B_j$ are unconstrained variables. 
If we place a single D3-brane at the singularity 
of the conifold, then we find
a $U(1)\times U(1)$ gauge theory coupled to fields $A_1, A_2$ with
charges $(1, -1)$ and $B_1, B_2$ with charges $(-1,1)$.

In constructing the generalization to the non-abelian theory on
$N$ D3-branes,
cancellation of the anomaly in the $U(1)$ R-symmetry requires that
the $A$'s and the $B$'s each have R-charge $1/2$. For consistency of
the duality it is necessary that we add
an exactly marginal superpotential which preserves the
$SU(2)\times SU(2)\times U(1)_R$ symmetry of the theory (this
superpotential produces a critical line related to the radius of 
$AdS_5\times T^{1,1}$). Since a marginal superpotential has R-charge
equal to 2 it must be quartic, and the symmetries fix it uniquely
up to overall normalization:
\be \label{superpotential}
W=\epsilon^{ij}
\epsilon^{kl}\tr A_iB_kA_jB_l\ .
\ee
Therefore, it was proposed in \cite{KW} that the $SU(N)\times
SU(N)$ SCFT with this superpotential is dual to type IIB strings
on $AdS_5\times T^{1,1}$.

This proposal can be checked in an interesting
way by comparing to a certain
$AdS_5\times {\bf S}^5/{\bf Z}_2$ background.  If ${\bf S}^5$ is described
by an equation
\be\label{beqnop}
\sum_{i=1}^6x_i^2=1,
\ee
with real variables $x_1,\dots, x_6$,
then the ${\bf Z}_2$ acts as $-1$ on four of the $x_i$ and
as $+1$ on the other two.  The importance of this choice is that this
particular
 ${\bf Z}_2$ orbifold of $AdS_5\times {\bf S}^5$
  has ${\cal N}=2$ superconformal symmetry.
Using orbifold results for D-branes \cite{dm}, this model has been 
identified \cite{ks} as an AdS dual of a $U(N)\times U(N)$ theory
with hypermultiplets transforming in
$(\N,\overline\N)\oplus (\overline \N,\N)$. From an ${\cal N}=1$ point of 
view, the hypermultiplets                
correspond to chiral multiplets $A_k,B_l$, $k,l=1,2$ in the 
$(\N,\overline \N)$ and $(\overline \N,\N)$  representations respectively.  
The model also contains, from an ${\cal N}=1$ point of view, chiral multiplets
$\Phi$ and $\tilde \Phi$ in the adjoint representations of the two
$U(N)$'s. 
The superpotential is
$$ g \Tr \Phi (A_1 B_1 + A_2 B_2) + g \Tr \tilde \Phi (B_1 A_1 + B_2 A_2)
\ .
$$
Now, let us add to the superpotential of this ${\bf Z}_2$ orbifold
a relevant term,
\be \label{relper}{m\over 2} (\Tr \Phi^2 - \Tr \tilde \Phi^2 )
\ .
\ee
It is straightforward to see what this does to the field theory.
We simply integrate out $\Phi$ and $\tilde \Phi$,
to find the superpotential
$$ {g^2\over m} \left [\Tr (A_1 B_1 A_2 B_2) - \Tr (B_1 A_1 B_2 A_2)
\right ]\ .
$$
This expression is the same as (\ref{superpotential}), so the
${\bf Z}_2$ orbifold with relevant perturbation (\ref{relper}) 
apparently flows to the $T^{1,1}$ model associated with the conifold.

Let us try to understand why this works from the point of view of the
geometry of ${\bf S}^5/{\bf Z}_2$.  The perturbation in (\ref{relper})
is odd under exchange of the two $U(N)$'s.  The exchange of the two $U(N)$'s
is the quantum symmetry of the $AdS_5\times {\bf S}^5/{\bf Z}_2$ orbifold
-- the symmetry that acts as $-1$ on string states in the twisted sector
and $+1$ in the untwisted  sector.  Therefore we
associate this perturbation with a twisted sector mode of string
theory on $AdS_5\times {\bf S}^5/\Z_2$.  The twisted sector mode
which is a relevant perturbation of the field theory is the blowup
of the orbifold singularity of ${\bf S}^5/\Z_2$ into the smooth space 
$T^{1,1}$. A somewhat different derivation of the field theory
on D3-branes at the conifold singularity, which is based on blowing
up a $\Z_2\times \Z_2$ orbifold, was given in \cite{MP}.

It is interesting to examine how various quantities change under
the RG flow from the ${\bf S}^5/{\bf Z}_2$ theory to the
$T^{1,1}$ theory. The behavior of the conformal anomaly (which is equal to
the $U(1)^3_R$ anomaly) was studied in \cite{Gubser}. Using the fact
that the chiral superfields carry
R-charge equal to $1/2$, on the field theory
side it was found that 
\be \label{anom}
{c_{IR}\over c_{UV}} = {27\over 32}
\ .
\ee
On the other hand, all 3-point functions calculated from supergravity on
$AdS_5\times X_5$ carry normalization factor inversely proportional
to ${\rm Vol}\ (X_5)$. Thus, on the supergravity side
\be
{c_{IR}\over c_{UV}} = {{\rm Vol}\ ({\bf S}^5/{\bf Z}_2)
\over {\rm Vol}\ (T^{1,1}) }
\ .
\ee
The volume of $T^{1,1}$ can be calculated from the exact formula for
the metric, (\ref{co}). One finds \cite{Gubser} 
\be 
\label{volume}
{\rm Vol}\ (T^{1,1})= {16 \pi^3\over 27} \ , 
\qquad {\rm Vol}\ ({\bf S}^5/{\bf Z}_2)= {\pi^3\over 2}\ ,
\ee
and the supergravity calculation
is in exact agreement with the field theory result 
(\ref{anom}) \cite{Gubser}.
This is a striking and highly sensitive test of the ${\cal N}=1$
dual pair constructed in \cite{KW,MP}.

\subsection{ Dimensions of Chiral Operators}

There is a number of further convincing checks of the duality between
this field theory and type IIB strings on $AdS_5\times T^{1,1}$.
Here we discuss the supergravity modes
which correspond to chiral primary operators. (For a more extensive
analysis of the spectrum of the model, see \cite{Ceres}.) 
For the $AdS_5\times {\bf S}^5$
case, these modes are mixtures of the conformal factors of the
$AdS_5$ and ${\bf S}^5$ and the 4-form field. 
The same has been shown to be true for the
$T^{1,1}$ case \cite{Gubser,RD,Ceres}. 
In fact, we may keep the discussion of such modes quite
general and consider $AdS_5\times X_5$ where $X_5$ is any Einstein manifold.

The diagonalization of such modes carried out in \cite{Kim}
for the ${\bf S}^5$ case is easily generalized to any $X_5$.
The mixing of the conformal factor and 4-form modes results in
the following mass-squared matrix,
\be m^2 = \pmatrix{ E+32 &  8E\cr  4/5 & E\cr}
\ee
where $E\geq 0$ is the eigenvalue of the Laplacian on $X_5$.
The eigenvalues of this matrix are
\be \label{masses}
 m^2 = 16 + E \pm 8 \sqrt{ 4+E}
\ .
\ee
We will be primarily interested in the modes which correspond
to picking the minus branch: they turn out to be the chiral primary
fields. For such modes there is a possibility of $m^2$ falling in
the range (\ref{range}) where there is a two-fold ambiguity in defining
the corresponding operator dimension. This happens for the eigenvalue $E$
such that
\be \label{bound}
5 \leq E \leq 21 \ .
\ee

First, let us recall the ${\bf S}^5$ case where the spherical
harmonics correspond to
traceless symmetric tensors of $SO(6)$, $d^{(k)}_{i_1\ldots i_k}$. 
Here $E= k(k+4)$, and it seems that
the bound (\ref{bound}) is satisfied for $k=1$. However, this is precisely the
special case where the corresponding mode is missing.
For $k=0$ there is no 4-form mode, hence no
mixing, while for $k=1$ one of the mixtures is the singleton \cite{Kim}.
Thus, all chiral primary 
operators in the ${\cal N}=4$ $SU(N)$ theory correspond to
the conventional branch of dimension, $\Delta_+$.
It is now well-known that this family of operators with dimensions
$\Delta= k$, $k=2,3,\ldots$
is $d^{(k)}_{i_1\ldots i_k} \Tr (X^{i_1} \ldots X^{i_k})$.
The absence of $k=1$ is related to the gauge group being $SU(N)$ rather
than $U(N)$. Thus, in this case we do not encounter operator
dimensions lower than $2$.

The situation is different for $T^{1,1}$. Here there is a family of wave
functions labeled by non-negative integer $k$, transforming under 
$SU(2)\times SU(2)$  as $(k/2,k/2)$, and with
$U(1)_R$ charge $k$ \cite{Gubser,RD,Ceres}.
The corresponding eigenvalues of the Laplacian
are 
\be \label{honeypot}
E(k)=3\left(k(k+2)
-{k^2\over 4}\right)\
.
\ee
In \cite{KW} it was argued that the dual chiral operators are
\be \label{chirop}
\tr (A_{i_1} B_{j_1} \ldots A_{i_k} B_{j_k} )
\ .
\ee
Since the F-term constraints in the gauge theory require that the
$i$ and the $j$ indices are separately symmetrized, we find that
their $SU(2)\times SU(2)\times U(1)$ quantum numbers agree with those
given by the supergravity analysis. In the field theory
the $A$'s and the $B$'s have dimension $3/4$, hence the dimensions
of the chiral operators are $3k/2$.

In studying the dimensions from the supergravity point of view, one encounters
an interesting subtlety discussed in section 5.2. While for $k>1$ only
the dimension $\Delta_+$ is admissible, for $k=1$ one could pick either 
branch. Indeed, from (\ref{honeypot}) we have
$E(1)=33/4$ which falls within the range (\ref{bound}). Here we
find that $\Delta_-=3/2$, while $\Delta_+=5/2$. Since the supersymmetry 
requires the corresponding dimension to be $3/2$, in this case
we have to pick the unconventional $\Delta_-$ branch. Choosing this
branch for $k=1$ and $\Delta_+$ for $k>1$ we indeed find
following \cite{Gubser,RD,Ceres}
that the supergravity analysis based on (\ref{masses}), (\ref{honeypot})
reproduces the dimensions $3k/2$ of the chiral operators (\ref{chirop}).
Thus, the conifold theory provides
a simple example of AdS/CFT duality where the $\Delta_-$ branch
has to be chosen for certain operators.

Let us also note that substituting $E(1)=33/4$ into (\ref{masses}) we find
$m^2=-15/4$ which corresponds to a conformally coupled scalar in $AdS_5$
\cite{Kim}. In fact, the short chiral
supermultiplet containing this scalar includes
another conformally coupled scalar and a massless fermion \cite{Ceres}. 
One of these
scalar fields corresponds to the lower component of the superfield
$\Tr (A_i B_j)$, which has dimension $3/2$, while the other
corresponds to the upper component which has dimension $5/2$. Thus,
the supersymmetry requires that we pick dimension $\Delta_+$ for one
of the conformally coupled scalars, and $\Delta_-$ for the other.

\subsection{Wrapped D3-branes as ``dibaryons''}

It is of further interest to consider various branes wrapped 
over the cycles of $T^{1,1}$ and attempt to 
identify these states in the field theory \cite{GK}.
For example, wrapped
D3-branes turn out to correspond to baryon-like operators $A^N$
and $B^N$ where the indices of both $SU(N)$ groups are fully
antisymmetrized.  For large $N$ the dimensions of such operators
calculated from the supergravity are found to be $3N/4$ \cite{GK}. 
This is in
complete agreement with the fact that the dimension of the chiral
superfields at the fixed point is $3/4$ and may be regarded as a direct
supergravity calculation of an anomalous dimension in the dual gauge
theory.

To show how this works in detail, we need to calculate the mass
of a D3-brane wrapped over a minimal volume 3-cycle.
An example of such a 3-cycle is 
the subspace at a constant value of $(\theta_2, \phi_2)$, and its volume
is found to be $V_3= 8\pi^2 L^3/9$ \cite{GK}.
The mass of the D3-brane
wrapped over the 3-cycle is, therefore,
 \be
m= V_3 {\sqrt\pi\over \kappa} = {8 \pi^{5/2} L^3\over 9\kappa}
\ .
\ee
For large $mL$, the corresponding operator dimension 
$\Delta$ approaches
\be 
m L= 
{8 \pi^{5/2} L^4\over 9\kappa} = {3\over 4} N
\ ,
\ee
where in the last step we used (\ref{quantiz}) with $\Omega_5$
replaced by ${\rm Vol}\ (T^{1,1})= 16 \pi^3/ 27$.

Let us construct the corresponding operators in the dual gauge theory.
Since the fields $A^{\alpha}_{k\beta}$, $k=1,2$,
carry an index $\alpha$ in the $\N$ of $SU(N)_1$ and an index $\beta$
in the $\overline{\N}$ of $SU(N)_2$,
we can construct color-singlet ``dibaryon'' operators
by antisymmetrizing completely with respect to both groups:
\be\label{BaryonOne}
{\cal B}_{1 l}= \epsilon_{\alpha_1 \ldots \alpha_N}  
\epsilon^{\beta_1\ldots \beta_N} D_{l}^{k_1\ldots k_N}  
\prod_{i=1}^N A^{\alpha_i}_{k_i  \beta_i} 
\ ,
\ee
where $D_l^{k_1\ldots k_N}$ is the completely 
symmetric $SU(2)$ Clebsch-Gordon
coefficient corresponding to forming the $N+1$ of $SU(2)$ out of $N$ 2's.
Thus the $SU(2)\times SU(2)$ quantum numbers of ${\mathcal B}_{1 l}$ are
$(N+1, 1)$. Similarly, we can construct ``dibaryon'' operators
which transform as $(1, N+1)$,
\be\label{BaryonTwo}
{\cal B}_{2 l}= \epsilon^{\alpha_1 \ldots \alpha_N}  
\epsilon_{\beta_1\ldots \beta_N} D_{l}^{k_1\ldots k_N}  
\prod_{i=1}^N B^{\beta_i}_{k_i  \alpha_i} 
\ .
\ee
Under the duality these operators map to D3-branes classically localized at 
a constant $(\theta_1,\phi_1)$. Thus, the existence of two types
of ``dibaryon'' operators is related on the supergravity
side to the fact that the base of the $U(1)$
bundle is ${\bf S}^2\times {\bf S}^2$. At the quantum level, the collective
coordinate for the wrapped D3-brane has to be quantized, and
this explains its $SU(2)\times SU(2)$ quantum numbers \cite{GK}.
The most basic check on the operator
identification is that, since the exact dimension of the $A$'s and the $B$'s
is $3/4$, the dimension of the ``dibaryon'' operators agrees exactly
with the supergravity calculation.

\subsection{Other ways of wrapping D-branes over cycles of $T^{1,1}$}

There are many other admissible ways of wrapping
branes over cycles of $T^{1,1}$ (for a complete list,
see \cite{Mukhi}). For example, a D3-brane may be
wrapped over a 2-cycle, which produces a string in $AdS_5$.
The tension of such a ``fat'' string
scales as 
$L^2/\kappa \sim N (g_s N)^{-1/2}/\alpha'$. The non-trivial
dependence of the tension on the 't~Hooft coupling $g_s N$ indicates
that such a string is not a BPS saturated object. This should be
contrasted with the tension of a BPS string obtained in \cite{Ed}
by wrapping a
D5-brane over ${\bf RP}^4$: $T\sim N/\alpha'$.

In discussing wrapped 5-branes, we will limit explicit statements
to D5-branes: since a $(p,q)$ 5-brane is an $SL(2,\Z)$
transform of a D5-brane, our discussion may be 
generalized to wrapped $(p,q)$ 5-branes using the $SL(2,\Z)$
symmetry of the Type IIB string theory.
If a D5-brane is wrapped over the entire $T^{1,1}$ then, according to
the arguments in \cite{Ed,GO}, it serves as a vertex connecting $N$
fundamental strings. Since each string ends on a charge in the
fundamental representation of one of the $SU(N)$'s, the resulting
field theory state is a baryon built out of external quarks.

If a
D5-brane is wrapped over an ${\bf S}^3$ then we find a membrane in
$AdS_5$. Although we have not succeeded in finding its field theoretic
interpretation, let us point out the following interesting effect.
Consider positioning a ``fat'' string made of a wrapped D3-brane
orthogonally to the membrane. As the string is brought through the
membrane, a fundamental string 
stretched between them 
is created. The origin of this effect
is creation of fundamental strings by crossing D5
and D3 branes, as shown in \cite{bdg,dfk}.

Finally, we discuss the very interesting case of D5-branes wrapped
over the 2-cycle, which have the right number of remaining dimensions
to be domain walls in $AdS_5$. The simplest domain wall is 
a D3-brane which is not wrapped over the compact manifold.
Through an analysis of the five-form flux carried over
directly from \cite{Ed} one can conclude that when one crosses the
domain wall, the effect in field theory is to change the gauge group
from $SU(N) \times SU(N)$ to $SU(N+1) \times SU(N+1)$.

The field theory interpretation of a D5-brane wrapped around ${\bf S}^2$ is
more interesting:  
if
on one side of the domain wall we have the original $SU(N) \times
SU(N)$ theory, then on the other side the theory is
$SU(N) \times SU(N+1)$ \cite{GK}.  
The matter fields $A_k$ and $B_k$ are still
bifundamentals, filling out $2 (\N,\overline{\N+1}) \oplus 2
(\overline{\N}, \N+1)$.  
One piece of evidence for this claim
is the way the D3-branes wrapped over the ${\bf S}^3$
behave when crossing the D5-brane domain wall.
In homology there is only one ${\bf S}^3$, but for
definiteness let us wrap the D3-brane around a particular three-sphere
${\bf S}^3_{(1)}$ which is invariant under the group $SU(2)_B$ under which
the fields $B_k$ transform.  The corresponding state in the $SU(N)
\times SU(N)$ field theory is ${\cal B}_1$ of \eno{BaryonTwo}.  In the
$SU(N) \times SU(N+1)$ theory, one has instead
  \eqn{TwoBaryons}{
   \epsilon_{\alpha_1 \ldots \alpha_N} \epsilon^{\beta_1 \ldots \beta_{N+1}} 
    A^{\alpha_1}_{\beta_1} \ldots A^{\alpha_N}_{\beta_N} 
      \qquad \hbox{or} \qquad
   \epsilon_{\alpha_1 \ldots \alpha_N} \epsilon^{\beta_1 \ldots \beta_{N+1}} 
    A^{\alpha_1}_{\beta_1} \ldots A^{\alpha_N}_{\beta_N} 
     A^{\alpha_{N+1}}_{\beta_{N+1}}
  }
 where we have omitted $SU(2)$ indices.  Either the upper index
$\beta_{N+1}$, indicating a fundamental of $SU(N+1)$, or the upper
index $\alpha_{N+1}$, indicating a fundamental of $SU(N)$, is free.

How can this be in supergravity?  The answer is simple: the wrapped
D3-brane must have a string attached to it. Indeed, after a wrapped
D3-brane has passed through the wrapped D5-brane domain wall, 
it emerges with a 
string attached to it due to the string creation by crossing D-branes
which together span 8 dimensions \cite{bdg,dfk}.

The domain wall in $AdS_5$ made out of $M$ wrapped D5-branes has the
following structure: on one side of it the 3-form field $H_{RR}$
vanishes, while on the other side there are $M$ units of flux of
$H_{RR}$ through the ${\bf S}^3$.  Thus, the supergravity dual of the
${\cal N}=1$ supersymmetric
$SU(N)\times SU(N+M)$ gauge theory involves adding $M$ units of RR 3-form
flux through the 3-cycle of $T^{1,1}$.
This theory is no longer conformal.  Instead, the relative gauge
coupling $g_1^{-2}-g_2^{-2}$ runs logarithmically, as pointed out in
\cite{KN} where the supergravity equations corresponding to this
situation were solved to leading order in $M/N$. The manifestation
of this RG flow in supergravity is the radial dependence of
the integral of $B_{NS}$ over the 2-cycle. This implies that the background
dual to the $SU(N)\times SU(N+M)$ theory has both the RR and NS-NS
3-form field strengths turned on.

Recently considerable progress in understanding such theories and their
gravity duals was made in \cite{KT,KS}. In particular, an exact
non-singular supergravity solution incorporating the 3-forms, the
5-form, and their back-reaction on the geometry has been 
constructed \cite{KS}.
This back-reaction turns the conifold into a deformed conifold
\be \label{defconi}
\sum_{a=1}^4 z_a^2 = \epsilon^2
\ ,
\ee
and introduces a ``warp factor'' so that the full 10-d geometry has
the form
\be \label{specans}
ds^2_{10} =   h^{-1/2}(r) (- dt^2 +  d\vec x^2) 
 +  h^{1/2}(r) d\tilde s_6^2 \ ,
\ee
where $d\tilde s_6^2$ is the metric of the deformed conifold.

Detailed discussion of this background is outside the scope of these notes.
We conclude by mentioning that it contains somewhat unexpected physics
in the UV: a cascade of Seiberg dualities \cite{NAD} 
which leads to jumps in the
rank of the gauge groups: $N\rightarrow N-M$ \cite{KT,KS}. 
Graceful exit from the cascade is achieved via the deformation of the
conifold. Near the apex, the warp factor $h^{1/2}$ approaches a
constant of order $g_s M$, which is the `t Hooft coupling of
the $SU(M)$ gauge theory found far in the IR. 
This behavior of the warp factor implies that the theory is
confining \cite{KS}. Moreover, the small $r$
part of the background incorporates a variety of other infrared phenomena
expected from this ${\cal N}=1$ supersymmetric
gauge theory: chiral symmetry breaking, 
glueball spectrum, baryons, domain walls separating inequivalent vacua, 
etc. For other recent work on relations between conifolds and
${\cal N}=1$ gauge theory, see \cite{MN,Vafa}.
We expect branes on the conifold to produce further insights into
gauge theory dynamics.

\section*{Acknowledgements}
I thank the organizers of TASI '99 for inviting me to present
these lectures in a very pleasant and stimulating atmosphere.
I am grateful to S. Gubser, N. Nekrasov, A. Peet, A. Polyakov, 
M. Strassler, W. Taylor, A. Tseytlin, M. Van Raamsdonk and
E. Witten, my collaborators on parts of the material reviewed in these
notes. I also thank C. Herzog and M. Krasnitz for reading the manuscript and
for useful comments on it.
This work  was supported in part by the NSF grant PHY-9802484 
and by the James S. McDonnell Foundation Grant No. 91-48.


\end{document}